\documentclass[11pt, oneside]{amsart}   	% use "amsart" instead of "article" for AMSLaTeX format
\usepackage{geometry}                		% See geometry.pdf to learn the layout options. There are lots.
\usepackage[parfill]{parskip}    		% Activate to begin paragraphs with an empty line rather than an indent
\usepackage{graphicx}				% Use pdf, png, jpg, or eps§ with pdflatex; use eps in DVI mode
\usepackage{amssymb}
\usepackage{amsmath}
\usepackage{amsfonts}
\usepackage{mathbbol}
\usepackage{amsaddr}
\numberwithin{equation}{section}

\title{A new perturbative expansion for fermionic functional integrals}
\author{Abhishek Goswami}
\address{Department of Mathematics, SUNY at Buffalo, Buffalo, NY 14260, USA}
\email{goswami3@buffalo.edu}

\begin{document}

\begin{abstract}
We construct a power series representation of the integrals of form
\begin{equation}
\text{log} \int d\mu_{S}(\psi, \bar{\psi})  \hspace{0.05 cm} e^{f(\psi, \bar{\psi}, \eta, \bar{\eta})}              \nonumber
\end{equation}
where $\psi, \bar{\psi}$ and $\eta, \bar{\eta}$ are Grassmann variables on a finite lattice in $d \geqslant 2$.
Our expansion has a local structure, is clean and provides an easy alternative to decoupling expansion
and Mayer-type cluster expansions in any analysis. As an example, we show exponential decay of 2-point 
truncated correlation function (uniform in volume) in massive Gross-Neveu model on a unit lattice.
\end{abstract}
\maketitle

\section{Introduction} 
Fermionic systems obey Pauli exclusion principle. For a fermionic quantum field theory with ultraviolet and infrared cut-offs,
a perturbative expansion of the logarithm of the generating functional generally converges since the higher powers of fermion
fields at a point are forbidden due to exclusion principle. This is not true for bosonic 
theories. Bosonic analysis often involves the use of heavy techniques such as Mayer cluster expansions as well as
decoupling expansion for the Gaussian measure  with covariance \textit{S}; $\mu_{S}$. There have been attempts to avoid 
Mayer-type cluster expansions for both bosonic and fermionic models.  
For example, Salmhofer and Wieczerkowski \cite{SW} express fermionic 
correlation functions as determinants of a matrix whose elements are given by the covariance and
show the convergence of perturbation theory. Their method does not work for bosonic theories.
Abdesselam and Rivasseau \cite{AR2} construct explicit fermionic tree expansions similar to that of bosons. 
While these methods eliminate Mayer-type expansions, a decoupling (interpolation) expansion for $\mu_{S}$ 
is still part of the analysis. The overall structure of these expansions is not conceptually straightforward. 

Decoupling expansions were pioneered by Glimm, Jaffe and Spencer \cite{GJS}.
The combinatorics was further simplified by Cammarota \cite{C} while proving the decay of correlation
functions in unbounded spin systems. More modern versions of decoupling expansions include
 Brydges and Kennedy \cite{BK} and Abdesselam and Rivasseau \cite{AR1}. 
 A key component of such expansions is the use of a weakening factor 
(for example, s-factor or w-factor) to carry out interpolation between given covariance matrix and a more local version of it.
However, this extra parameter further complicates the analysis, for example, the differential aspects with respect to the 
weakening factor and other combinatorics must also be studied. 
In this paper we present a new expansion which completely bypasses all decoupling expansions as well as standard
Mayer cluster expansions. Our expansion is conceptually simple and does not involve any artificial weakening factor 
while avoiding the standard polymer construction associated with Mayer-type expansions. 

The method we present here originated in our earlier work on the Higgs mechanism  \cite{G}.
We proved the existence of a mass gap in a weakly coupled U(1) Higgs theory
on a unit lattice. We applied a power series cluster expansion due to Balaban, Feldman, Kn$\ddot{\text{o}}$rrer and Trubowitz
\cite{BFKT} designed for bosonic functional integrals having Gaussian measure with unit covariance (decoupled).   
Such expansion for fermions is not in the literature. Here we construct a similar power series expansion for fermionic
functional integrals but with a more general (non local) covariance.
The two expansions together can be used to study systems involving both fermions and bosons.
Indeed this is one of our motivations for this work. 

The idea is to (i) make a change of variables $\psi \rightarrow S \psi$ and transform a Gaussian measure
$\mu_{S}$ into a one with unit covariance; $\mu_{I}$, (ii) write a power series for $f(S \psi, \bar{\psi}, \eta, \bar{\eta})$
and (iii) given a series for $f(S \psi, \bar{\psi}, \eta, \bar{\eta})$ generate a series expansion for the log of
generating functional.
Our analysis works in $d \geqslant 2$ for any finite lattice of arbitrary spacing but for the purpose of clarity we work on a unit lattice.
As an application we show exponential decay of fermion-fermion truncated correlation function in massive Gross-Neveu model
on a unit lattice. For other methods of fermionic cluster expansions we refer reader to 
Feldman, Magnen, Rivasseau and Trubowitz \cite{FMRT} and Dimock \cite{D}.  

\subsection{Fermion field} Let $\Lambda \subset \mathbb{Z}^{d}$ and $x, y \in \Lambda$. Let \textit{V} be a vector space over 
field $\mathbb{C}$ with basis $\psi_{\alpha,a}(x)$ and $\bar{\psi}_{\alpha,a}(x)$ where $\alpha$ is spinor index and
$a$ is the index of some internal symmetry of the fermions. 
Let $V^{\prime}$ be a vector space over field $\mathbb{C}$ with basis $\eta_{\alpha,a}(y)$ and $\bar{\eta}_{\alpha,a}(y)$. 
$\psi_{\alpha,a}(x), \bar{\psi}_{\alpha,a}(x)$ and $\eta_{\alpha,a}(y), \bar{\eta}_{\alpha,a}(y)$ are taken to be
Grassmann variables satisfying anticommuting property
\begin{equation}
\begin{aligned}
&\{\psi_{\alpha,a}(x), \bar{\psi}_{\alpha,a}(x)\} = 0, \hspace{0.6 cm} 
\{\eta_{\alpha,a}(y), \bar{\eta}_{\alpha,a}(y)\} = 0, \hspace{0.6 cm} \{\psi_{\alpha,a}(x), \eta_{\alpha,a}(x)\} = 0, \\
& \{\bar{\psi}_{\alpha,a}(x), \bar{\eta}_{\alpha,a}(x)\} = 0, \hspace{0.6 cm}
\{\bar{\psi}_{\alpha,a}(x), \eta_{\alpha,a}(x)\} = 0, \hspace{0.6 cm} \{\psi_{\alpha,a}(x), \bar{\eta}_{\alpha,a}(x)\} = 0. 
\end{aligned}
\end{equation}
We consider Grassmann algebra generated by $\psi_{\alpha,a}(x), \bar{\psi}_{\alpha,a}(x)$ and 
$\eta_{\alpha,a}(y), \bar{\eta}_{\alpha,a}(y)$.
Let $\xi$ stands for $(x, \alpha, a, \omega)$ and $z$ stands for $(y, \alpha, a, \omega)$ with $\omega = (0,1)$.
Define
\begin{equation}
 \psi(\xi) = \begin{cases}
 \psi_{\alpha,a}(x)      & \text{if} \hspace{0.4 cm} \xi = (x, \alpha, a, 0) \\
\bar{\psi}_{\alpha,a}(x)   & \text{if} \hspace{0.4 cm} \xi = (x, \alpha, a, 1)
\end{cases}
\end{equation} 
\begin{equation}
 \eta(z) = \begin{cases}
 \eta_{\alpha,a}(y)      & \text{if} \hspace{0.4 cm} z = (y, \alpha, a, 0) \\
\bar{\eta}_{\alpha,a}(y)   & \text{if} \hspace{0.4 cm} z = (y, \alpha, a, 1)
\end{cases}
\end{equation} 
Then  
$\psi(\xi) \psi(\xi^{\prime}) = - \psi(\xi^{\prime}) \psi(\xi)$ and $ \eta(z) \eta(z^{\prime}) = - \eta(z^{\prime}) \eta(z)$ generate  
the algebra. 

Grassmann algebra is the vector space $\mathcal{V}$ defined as
\begin{equation}
\mathcal{V} = \bigoplus_{k= 0}^{\text{dim} (V \oplus V^{\prime})}  \bigwedge\nolimits^{k} V \oplus V^{\prime}
\end{equation}
where $\bigwedge\nolimits^{0} V \oplus V^{\prime} = \mathbb{C}$ and $\bigwedge\nolimits^{k} V \oplus V^{\prime}$ 
denotes \textit{k} fold antisymmetric tensor
product of $ V \oplus V^{\prime}$  with itself and $\text{dim} (V \oplus V^{\prime})$ is the dimension of the
vector space $V \oplus V^{\prime}$. An element of vector space $\mathcal{V}$ is given by
\begin{equation}
\begin{aligned}
f(\psi, \eta)  &= \sum_{n,m\geqslant0}  \sum_{\substack{\xi_{1},\cdots, \xi_{n} \\ z_{1},\cdots, z_{m}}} 
a(\xi_{1},\cdots, \xi_{n}, z_{1},\cdots, z_{m})\psi(\xi_{1}) \cdots \psi(\xi_{n}) \eta (z_{1}) \cdots \eta(z_{m}) 
\end{aligned}
\end{equation}
where the coefficients $a(\xi_{1},\cdots, \xi_{n}, z_{1},\cdots, z_{m}) \in \mathbb{C}$. $f(\psi, \eta)$ is known as Grassmann  
element with kernel $a(\xi_{1},\cdots, \xi_{n}, z_{1},\cdots, z_{m})$. The coefficients (or kernel)
are not assumed to be antisymmetric. But they can taken to be antisymmetric without changing $f$.
  
\subsection{Power series} A power series of Grassmann variables is an element of the Grassmann algebra.
It is an element of the vector space $\mathcal{V}$ as (1.5).  
For example, let $X \subset \Lambda$ and
\begin{equation}
\begin{aligned}
f(\psi) &= \sum_{x \in X} \sum_{\alpha,\beta,a,b} 
 \bar{\psi}_{\alpha,a}(x) \bar{\psi}_{\beta,b}(x) \psi_{\alpha,a}(x)  \psi_{\beta,b}(x) \\
 &= \sum_{x_{1}, x_{2}, x_{3}, x_{4} \in X} \sum_{\alpha,\beta,a,b} 
 a_{\alpha,\beta,a,b} (x_{1}, x_{2}, x_{3}, x_{4}) \hspace{0.05 cm}
  \bar{\psi}_{\alpha,a}(x_{1}) \bar{\psi}_{\beta,b}(x_{2}) \psi_{\alpha,a}(x_{3}) \psi_{\beta,b}(x_{4})
\end{aligned}
\end{equation}
where $a_{\alpha,\beta,a,b} (x_{1}, x_{2}, x_{3}, x_{4}) =  
\delta_{\alpha,\beta,a,b}(x_{1} - x_{2}) \delta_{\beta,\alpha,b,a}(x_{2} - x_{3}) \delta_{\alpha,\beta,a,b}(x_{3} - x_{4})$. 
Similarly for
\begin{equation}
\begin{aligned}
f(\psi, \eta) &= \sum_{x \in X} \sum_{\alpha,a} \hspace{0.1 cm}
[ \bar{\psi}_{\alpha,a}(x) \eta_{\alpha,a}(x) + \bar{\eta}_{\alpha,a}(x) \psi_{\alpha,a}(x)] \\
&= \sum_{x_{1}, x_{2} \in X} \sum_{\alpha,a}
a_{\alpha,a}(x_{1}, x_{2}) \hspace{0.05 cm}[\bar{\psi}_{\alpha,a}(x_{1}) \eta_{\alpha,a}(x_{2}) + 
\bar{\eta}_{\alpha,a}(x_{1}) \psi_{\alpha,a}(x_{2})]
\end{aligned}
\end{equation}
where $a_{\alpha,a}(x_{1}, x_{2}) = \delta_{\alpha,a}(x_{1} - x_{2})$. 
The series coefficients in (1.6) and (1.7) are not antisymmetric.  

\textbf{Remark 1}. In Euclidean fermionic theory correlation functions are integrals of such elements of
Grassmann algebra. In a renormalization group approach,
\begin{center}
$\psi(\xi)$ - fluctuation field and
$ \eta(z)$ - remaining variables.
\end{center}
 
\textbf{Notation}. Define a n-component vector $\vec{\xi}$ as  
\begin{equation}
\vec{\xi} = \{\xi_{1}, \xi_{2}, \cdots, \xi_{n} \}.
\end{equation}
Define concatenation of two vectors $\vec{\xi}$ and $\vec{z}$ as
\begin{equation}
\begin{aligned}
\vec{\xi} \circ \vec{z} &= \{\xi_{1}, \xi_{2}, \cdots, \xi_{n}, z_{1}, \cdots, z_{m}\} \\
(\vec{\xi}_{1},\cdots, \vec{\xi}_{s}) &\circ (\vec{z}_{1},\cdots, \vec{z}_{s}) =
(\vec{\xi}_{1} \circ \vec{z}_{1}, \cdots, \vec{\xi}_{s} \circ \vec{z}_{s}).
\end{aligned}
\end{equation}
For a n-component vector $\vec{\xi}$, write
\begin{equation}
\psi (\vec{\xi}) = \psi (\xi_{1}) \cdots \psi (\xi_{n}).
\end{equation}
Write an element of the Grassmann algebra  in a power series representation as
\begin{equation}
\begin{aligned}
f(\psi, \eta)  &= \sum_{n,m\geqslant0}  \sum_{\substack{\xi_{1},\cdots, \xi_{n} \\ z_{1},\cdots, z_{m}}} 
a(\xi_{1},\cdots, \xi_{n}, z_{1},\cdots, z_{m})\psi(\xi_{1}) \cdots \psi(\xi_{n}) \eta (z_{1}) \cdots \eta(z_{m}) \\
&= \sum_{n,m \geqslant 0} \sum_{\substack{\vec{\xi} = \{\xi_{1},\cdots, \xi_{n}\} \\ 
\vec{z} = \{z_{1},\cdots, z_{m} \} }} a(\vec{\xi}, \vec{z}) \hspace{0.05 cm} \psi (\vec{\xi}) \eta(\vec{z}).
\end{aligned}
\end{equation}
Any change in the order of fields is accompanied by a negative sign.  
Let
\begin{equation}
a_{\eta}(\vec{\xi}) = \sum_{\vec{z}} a(\vec{\xi}, \vec{z}) \hspace{0.05 cm}  \eta(\vec{z})
\end{equation}
and rewrite (1.11) as
\begin{equation}
f(\psi, \eta) = \sum_{\vec{\xi}} a_{\eta}(\vec{\xi}) \hspace{0.05 cm} \psi (\vec{\xi}). 
\end{equation}
The power series representation of a function of Grassmann element $e^{f(\psi, \eta)}$ is given by
\begin{equation}
e^{f(\psi, \eta)} = \sum_{l=0}^{\infty} \frac{1}{l!} f(\psi, \eta)^{l} = 1 +  \sum_{l=1}^{\infty}  \frac{1}{l!} 
 \sum_{\vec{\xi}_{1},\cdots, \vec{\xi}_{l} \in X} (-1)^{\#} a_{\eta}(\vec{\xi}_{1}) \cdots  a_{\eta}(\vec{\xi}_{l}) 
 \hspace{0.05 cm} \psi (\vec{\xi}_{1}) \cdots  \psi (\vec{\xi}_{l}).
\end{equation}
where $\#$ denotes the number of interchanges of $\psi, \eta$.
Define a normalized Gaussian measure with unit covariance as
\begin{equation}
d\mu_{I}(\psi) = \frac{\prod_{\alpha,a,x} d\bar{\psi}_{\alpha,a}(x) d\psi_{\alpha,a}(x) \hspace{0.05 cm}
e^{- \langle \bar{\psi}, \psi \rangle}}
{\int \prod_{\alpha,a,x} d\bar{\psi}_{\alpha,a}(x) d\psi_{\alpha,a}(x) \hspace{0.05 cm}
e^{- \langle \bar{\psi}, \psi \rangle}}
\end{equation}
A general Gaussian measure can be put in this form by a simple change of variables. 

Denote $\Xi(\eta) = \int d\mu_{I}(\psi) \hspace{0.05 cm} e^{f(\psi, \eta)}$. A power series 
expansion of $\text{log}\hspace{0.1 cm} \Xi(\eta)$ is then given by
\begin{equation}
\begin{aligned}
\text{log}\hspace{0.1 cm} \Xi(\eta) &= b_{0} +
 \eta(z_{1}) \hspace{0.05 cm} b(z_{1}, z_{2}) \hspace{0.05 cm}  \eta(z_{2}) +
 \eta(z_{1}) \eta(z_{2}) \hspace{0.05 cm} b(z_{1}, z_{2}, z_{3}, z_{4}) \hspace{0.05 cm} \eta(z_{3}) \eta(z_{4})
 +  \cdots \\
&= \sum_{m \geqslant 0} \eta(z_{1}) \cdots \eta(z_{m}) \hspace{0.05 cm} b(z_{1}, \cdots, z_{2m}) 
\hspace{0.05 cm}  \eta(z_{m+1}) \cdots \eta(z_{2m}) \\
&=  \sum_{m \geqslant 0} \eta(\vec{z}_{m}) \hspace{0.05 cm} b(\vec{z}) \hspace{0.05 cm} \eta(\vec{z}_{2m}) 
\end{aligned}
\end{equation}
where the coefficient system $b(z_{1}, \cdots, z_{2m})$ is also not assumed to be antisymmetric. 
This is the desired expansion.  

\textbf{Weight system.} For a n-component vector $\vec{\xi}$ a weight system $w(\vec{\xi})$ is a function
which assigns a positive number $w(\xi_{1}, \cdots, \xi_{n})$ to $\vec{\xi}$ and satisfies
the following two properties:
\begin{enumerate}
  \item $w(\xi_{1}, \cdots, \xi_{n})$ is invariant under the permutations of the components
  of $\vec{\xi}$ and
  \item for any two vectors $\vec{\xi}$ and $\vec{z}$ with 
  $\text{supp}(\vec{\xi}) \cap  \text{supp}(\vec{z}) \neq \oslash $
  \begin{equation}
w(\vec{\xi}, \vec{z}) = w(\vec{\xi} \circ \vec{z}) \leqslant w(\vec{\xi}) \hspace{0.05 cm} w(\vec{z}).      \nonumber
\end{equation}
\end{enumerate}
Below we discuss one example of a weight system. For more examples we refer reader to
\cite{BFKT}. 

\textbf{Norm.}
Let $w_{h_{1}, h_{2}}(\vec{\xi}, \vec{z}) =  e^{\kappa t(\text{supp}(\vec{\xi}, \vec{z}))} h_{1}^{n} h_{2}^{m}$ 
be the weight system with mass $\kappa$ giving weight at least $h_{1}$ to $\psi$ and $h_{2}$ to $\eta$.
Define 
\begin{equation}
 |a|_{w_{h_{1}, h_{2}}} = \sum_{ n, m \geqslant 0}
\max\limits_{\xi \in X}  \hspace{0.05 cm} \max\limits_{\substack{1\leqslant i \leqslant n \\ 1\leqslant j \leqslant m}} 
\sum_{\substack{\xi_{1}, \cdots, \xi_{n}, z_{1}, \cdots, z_{m} \in X \\ \xi_{i} = \xi \hspace{0.05 cm} or \hspace{0.05 cm}
z_{j} = \xi}}  e^{\kappa t(\text{supp}(\vec{\xi}, \vec{z}))} 
 h_{1}^{n} h_{2}^{m} \hspace{0.05 cm}  |a(\xi_{1}, \cdots, \xi_{n}, z_{1}, \cdots, z_{m})|
 \end{equation}
where $a(\xi_{1}, \cdots, \xi_{n}, z_{1}, \cdots, z_{m})$ is the antisymmetric realization of $a$.
Then the norm is defined as  $||f||_{w_{h_{1}, h_{2}}} = |a|_{w_{h_{1}, h_{2}}}$. Here $\max\limits_{\xi \in X} $ 
breaks the translation invariance.

For example, for $f(\psi) = \sum_{x \in X} \sum_{\alpha,\beta,a,b} 
 \bar{\psi}_{\alpha,a}(x) \bar{\psi}_{\beta,b}(x) \psi_{\alpha,a}(x)  \psi_{\beta,b}(x)$ as in (1.6),
the norm  $||f||_{w_{h_{1}, h_{2}}} = d^{2} N^{2} h_{1}^{4}$ and for 
$f(\psi, \eta) =   \sum_{x \in X} \sum_{\alpha,a} \hspace{0.1 cm}
[ \bar{\psi}_{\alpha,a}(x) \eta_{\alpha,a}(x) + \bar{\eta}_{\alpha,a}(x) \psi_{\alpha,a}(x)]$ 
as in (1.7), norm $||f||_{w_{h_{1}, h_{2}}} = 2 d N h_{1} h_{2}$, where $d$ is the dimension of the lattice and
$N$ is the number of internal symmetry of fermions. Note that because of the delta decay in these
examples $\text{supp}(\vec{\xi}, \vec{z})$ only has a single point due to which the length of the tree
$t(\text{supp}(\vec{\xi}, \vec{z}))$ is zero.

The norm for $\text{log}\hspace{0.1 cm} \Xi(\eta)$ is defined in the same manner. 
Let $w_{h}(\vec{z}) = e^{\kappa t(\text{supp}(\vec{z}))} h^{n(\vec{z})}$ be the weight system of mass 
$\kappa$ (where  $n(\vec{z})$ is the $\#$ of sites in $\vec{z}$) giving weight $h$ to the field $\eta$. Define
\begin{equation}
|b|_{w_{h}} = \sum_{n \geqslant 0}   
\max\limits_{z \in X} \max\limits_{1\leqslant i \leqslant n} \sum_{\substack{(z_{1}, \cdots, z_{n}) \in X 
 \\ z_{i} = z}} w_{h}(\vec{z}) |b (z_{1}, \cdots, z_{n})|
\end{equation}
Then the norm $||\text{log}\hspace{0.1 cm} \Xi(\eta)||_{w_{h}} = |b|_{w_{h}}$.

\textbf{Theorem 1} Let $f(\psi, \eta)$ has a power series representation with
coefficient system $a (\vec{\xi}, \vec{z})$. 
Then there exists a coefficient system $b(\vec{z})$ having same form   
as $a (\vec{\xi}, \vec{z})$ such that 
$\text{log}\hspace{0.1 cm} \Xi(\eta) = \sum_{\vec{z}} b (\vec{z}) \eta (\vec{z})$.
Denote $\text{log}\hspace{0.1 cm} \Xi(\eta) = H(\eta)$ and
set $||H||_{w_{h}}=|b|_{w_{h}}$, then if $||f||_{w_{4, h}} < \frac{1}{16}$ 
\begin{equation}
||H - H(0)||_{w_{h}} \leqslant 
\frac{||f||_{w_{4, h}}}{1 - 16 ||f||_{w_{4,h}}}. 
\end{equation}
 
\textbf{Remark 2}. The series expansion (1.16) converges in norm when $|b|_{w_{h}} < \infty$.

\section{Proof}
The algebra and combinatorics of the proof follows that of the bosonic case due to 
Balaban, Feldman, Kn$\ddot{\text{o}}$rrer and Trubowitz \cite{BFKT}. 

\textbf{Definition.} Let $X \subset \Lambda$ and let $\vec{x}_{1}, \cdots, \vec{x}_{l}$ be a collection of vectors 
in $X$. Denote $X_{i} = \text{supp}(\vec{x}_{i})$ for $i = 1\hspace{0.1 cm} \text{to} \hspace{0.1 cm}l$.
Consider the set $\{X_{1}, \cdots, X_{l}\}$. Define \textit{incidence graph} $G(X_{1}, \cdots, X_{l})$ 
to be a graph with vertices $\{1, \cdots, l\}$ and edges $(i,j)$ whenever $X_{i} \cap X_{j} \neq \oslash$.
The set $\{\vec{x}_{1}, \cdots, \vec{x}_{l}\}$ is called \textit{connected} if the incidence graph  $G(X_{1}, \cdots, X_{l})$ 
is connected. For any set $Z \subset X$ define a \textit{connected cover} $\mathcal{C}(Z)$ 
to be the set of all ordered connected \textit{l}-tuples $\{\vec{x}_{1}, \cdots, \vec{x}_{l}\}$ for which
$Z = X_{1} \cup \cdots \cup X_{l}$.

\textbf{Proposition 1} Let $a(\vec{\xi}, \vec{z})$ be a coefficient system  
representing a power series of $f(\psi,\eta)$ as in (1.11), (1.13). 
Let $Z = \text{supp} (\vec{\xi}_{1}, \cdots, \vec{\xi}_{l})$ as in (1.14). Let $ \mathcal{C}(Z)$ denote a connected
cover of $Z$. Then there exists a function $K_{0}(Z, \eta)$ given by
\begin{equation}
K_{0}(Z, \eta) = (\pm 1) \sum_{k=1}^{\infty} \frac{1}{k!}  \sum_{(\vec{\xi}_{1}, \cdots, \vec{\xi}_{k}) \in  \mathcal{C}(Z)}
 a_{\eta}(\vec{\xi}_{1}) \cdots  a_{\eta}(\vec{\xi}_{k})
\end{equation}
such that
\begin{equation}
\int d\mu_{I}(\psi) \hspace{0.05 cm} e^{f(\psi, \eta)} = 1 +  \sum_{n=1}^{\infty} \frac{1}{n!} 
\sum_{\substack{Z_{1}, \cdots, Z_{n} \subset X \\ Z_{i} \cap Z_{j} = \oslash}} (-1)^{\#} \prod_{j=1}^{n} K_{0}(Z_{j}, \eta).
\end{equation}

\textit{Proof} First start with Eq. (1.14)
\begin{equation}
e^{f(\psi, \eta)} =  1 +  \sum_{l=1}^{\infty}  \frac{1}{l!}
 \sum_{\vec{\xi}_{1},\cdots, \vec{\xi}_{l} \in X} (-1)^{\#} a_{\eta}(\vec{\xi}_{1}) \cdots  a_{\eta}(\vec{\xi}_{l}) 
 \hspace{0.05 cm} \psi (\vec{\xi}_{1}) \cdots  \psi (\vec{\xi}_{l}). \nonumber
\end{equation}
It is important that the $\text{supp}(\vec{\xi}_{i})$ stays close (overlaps) to the $\text{supp}(\vec{\xi}_{j})$ for $i \neq j$
so that the resulting coefficient system $b(\vec{z})$ also have a decay similar to that of $a(\vec{\xi}, \vec{z})$.
Recall that $\xi = (x, \alpha, a, \omega)$. Here \textit{overlap} refers to the overlap of sites only.  
We identify the sites of $\vec{\xi}_{i}$ as $\vec{x}_{i}$ for $i = 1\hspace{0.1 cm} \text{to} \hspace{0.1 cm}l$.
Let $X_{i} = \text{supp}(\vec{\xi}_{i})$ and $Z = X_{1} \cup \cdots \cup X_{l}$.

Next we decompose $Z$ into pairwise disjoint subsets
$\{Z_{1}, \cdots, Z_{n}\}$.  Divide $\{1, \cdots, l\}$ into pairwise disjoint subsets $I_{1}, \cdots, I_{n}$
such that for each $1 \leqslant j \leqslant n$, $\vec{\xi}_{i}, i \in I_{j}$ is $\mathcal{C}(Z_{j})$. 
Following the combinatorics in \cite{BFKT} rewrite (1.14) as
\begin{equation}
\begin{aligned}
 \int d\mu_{I}(\psi) \hspace{0.05 cm} e^{f(\psi, \eta)} &= 1 +  \sum_{n=1}^{\infty} \frac{1}{n!} 
\sum_{\substack{Z_{1}, \cdots, Z_{n} \subset X \\ Z_{i} \cap Z_{j} = \oslash}}  \sum_{k_{1}, \cdots, k_{n} \geqslant 1}  
(-1)^{\#} \int d\mu_{I}(\psi)\\ 
&  \prod_{j=1}^{n}  \frac{1}{k_{j}!}  \sum_{(\vec{\xi}_{1}, \cdots, \vec{\xi}_{k_{j}}) \in  \mathcal{C}(Z_{j})}
 a_{\eta}(\vec{\xi}_{1,j}) \cdots  a_{\eta}(\vec{\xi}_{k_{j},j})
\hspace{0.05 cm} \psi (\vec{\xi}_{1,j}) \cdots  \psi (\vec{\xi}_{k_{j},j}).
\end{aligned}
\end{equation}
Since $Z_{i} \cap Z_{j} = \oslash$, the measure factorizes
\begin{equation}
 \int d\mu_{I}(\psi) \prod_{j=1}^{n}  \psi (\vec{\xi}_{1,j}) \cdots  \psi (\vec{\xi}_{k_{j},j})
 = \prod_{j=1}^{n} \int d\mu_{I}(\psi) \psi (\vec{\xi}_{1,j}) \cdots  \psi (\vec{\xi}_{k_{j},j}). 
\end{equation}
Note that for every $\vec{\xi}_{i,j}$ (assuming there are \textit{n} sites in $\vec{\xi}_{i,j}$)
\begin{equation}
\begin{aligned}
\int d\mu_{I}(\psi) \hspace{0.05 cm} & \bar{\psi}_{\beta_{1},b_{1}}(x_{1}) \cdots \bar{\psi}_{\beta_{n},b_{n}}(x_{n})
\psi_{\alpha_{1},a_{1}}(y_{1}) \cdots \psi_{\alpha_{m},a_{m}}(y_{m}) = \\ 
&\begin{cases}
\pm 1    & \text{if} \hspace{0.4 cm} n =m \hspace{0.2 cm} \text{and} \hspace{0.2 cm} 
\{(\beta_{i}, b_{i}, x_{i}) \} = \{(\alpha_{i}, a_{i}, y_{i})\} \\
0   & \text{otherwise}.
\end{cases}
\end{aligned}
\end{equation}
For every $Z \in \{Z_{1},\cdots, Z_{n}\}$, we now identify $K_{0}(Z, \eta)$  
such that
\begin{equation}
\int d\mu_{I}(\psi) \hspace{0.05 cm} e^{f(\psi, \eta)} = 1 +  \sum_{n=1}^{\infty} \frac{1}{n!} 
\sum_{\substack{Z_{1}, \cdots, Z_{n} \subset X \\ Z_{i} \cap Z_{j} = \oslash}} (-1)^{\#} \prod_{j=1}^{n} K_{0}(Z_{j}, \eta). \nonumber
\end{equation}
This completes the proof of the proposition.

\textbf{Remark 3}. The integral (2.5) forces every site to have even number of fields; $\psi$ and $\bar{\psi}$ and 
thus, $\eta$ and $\bar{\eta}$.  

To take care of the pairwise disjoint condition of $Z_{j}'s$, follow the standard procedure and by standard argument \cite{BFKT},
\begin{equation}
\text{log}\hspace{0.1 cm} \int d\mu_{I}(\psi) \hspace{0.05 cm} e^{f(\psi, \eta)} = 
 \sum_{n=1}^{\infty} \frac{1}{n!} 
\sum_{Z_{1}, \cdots, Z_{n} \subset X } \rho^{T}(Z_{1}, \cdots, Z_{n}) \hspace{0.05 cm}
(-1)^{\#} \prod_{j=1}^{n} K_{0}(Z_{j}, \eta)
\end{equation}
where $\rho^{T}$ is a certain function with a property
$\rho^{T}(Z_{1}, \cdots, Z_{n}) = 0$ if $\{Z_{j}\}$ has at least one disjoint pair.

\textbf{Proposition 2} Let $a(\vec{\xi}, \vec{z})$ be a coefficient system and 
$K_{0}(Z, \eta)$ be the function as defined.   
Then there exists a coefficient system
$b( \vec{z})$ such that $\text{log}\hspace{0.1 cm}  \int d\mu_{I}(\psi) \hspace{0.05 cm} e^{f(\psi, \eta)}
 = \sum_{\vec{z} \in X} b (\vec{z}) \eta (\vec{z})$.

\textit{Proof} Using (1.12) ($a_{\eta}(\vec{\xi}) = \sum_{\vec{z}} a(\vec{\xi}, \vec{z}) \hspace{0.05 cm}  \eta(\vec{z})$)
rewrite $K_{0}(Z, \eta)$ as
\begin{equation}
\begin{aligned}
K_{0}(Z, \eta) &= (\pm 1) \sum_{k=1}^{\infty} \frac{1}{k!} 
\sum_{\substack{(\vec{\xi}_{1}, \cdots, \vec{\xi}_{k}) \in  \mathcal{C}(\text{supp} \hspace{0.5 mm} \vec{\xi}) \\
\vec{\xi}_{1}\circ \cdots \circ \vec{\xi}_{k} = \vec{\xi}}} \sum_{\substack{(\vec{z}_{1}, \cdots, \vec{z}_{k}) \\
\vec{z}_{1}\circ \cdots \circ \vec{z}_{k} = \vec{z}}} a(\vec{\xi}_{1}, \vec{z}_{1}) \cdots a(\vec{\xi}_{k}, \vec{z}_{k})
\hspace{0.1 cm} \eta(\vec{z}) \\
&= \sum_{\substack{\vec{\xi} \in X, \vec{z} \in X \\ \text{supp} \vec{\xi} = Z}} 
(\pm 1) \sum_{k=1}^{\infty} \frac{1}{k!} 
\sum_{\substack{(\vec{\xi}_{1}, \cdots, \vec{\xi}_{k}) \in  \mathcal{C}(\text{supp} \hspace{0.5 mm} \vec{\xi}) \\
\vec{\xi}_{1}\circ \cdots \circ \vec{\xi}_{k} = \vec{\xi}}} \sum_{\substack{(\vec{z}_{1}, \cdots, \vec{z}_{k}) \\
\vec{z}_{1}\circ \cdots \circ \vec{z}_{k} = \vec{z}}} a(\vec{\xi}_{1}, \vec{z}_{1}) \cdots a(\vec{\xi}_{k}, \vec{z}_{k})
\hspace{0.1 cm} \eta(\vec{z}) \\
&= \sum_{\substack{\vec{\xi} \in X, \vec{z} \in X \\ \text{supp} \vec{\xi} = Z}} 
\tilde{a}(\vec{\xi}, \vec{z}) \hspace{0.05 cm} \eta(\vec{z}).
\end{aligned}
\end{equation}
where first we have grouped vectors together as concatenation and then we sum over concatenations.
Following (2.6) and (2.7) define
\begin{equation}
b(\vec{z}) =  \sum_{n=1}^{\infty} \frac{1}{n!} \sum_{\substack{(\vec{z}_{1}, \cdots, \vec{z}_{n}) \\
\vec{z}_{1}\circ \cdots \circ \vec{z}_{n} = \vec{z}}} \sum_{\vec{\xi}_{1}, \cdots, \vec{\xi}_{n} \in X}
(-1)^{\#} \rho^{T}(\text{supp} \vec{\xi}_{1}, \cdots, \text{supp} \vec{\xi}_{n})
\prod_{j=1}^{n} \tilde{a}(\vec{\xi}_{j}, \vec{z}_{j})
\end{equation}
where $ \rho^{T}(\text{supp} \vec{\xi}_{1}, \cdots, \text{supp} \vec{\xi}_{n}) = 0$ if 
$G(\text{supp} \vec{\xi}_{1}, \cdots, \text{supp} \vec{\xi}_{n})$ is not connected.  
Then (2.6) becomes $\text{log}\hspace{0.1 cm}  \int d\mu_{I}(\psi) \hspace{0.05 cm} e^{f(\psi, \eta)}
= H(\eta) = \sum_{\vec{z} \in X} b (\vec{z}) \eta (\vec{z})$.  
This completes the proof of proposition. 

Note that the coefficient system $b (\vec{z})$ depends on $ \tilde{a}(\vec{\xi}, \vec{z})$ which in
turn depends on $a(\vec{\xi}, \vec{z})$. Thus, $b (\vec{z})$ has the same form as the coefficient system $a(\vec{\xi}, \vec{z})$.

\subsection{Estimate on norms} 
First rewrite  coefficient systems  as \cite{BFKT},
\begin{equation}
b (\vec{z}) =  \sum_{n=1}^{\infty} \frac{1}{n!} \sum_{\substack{T, \text{labelled tree with} \\ \text{vertices} 1,\cdots, n}}
\sum_{\vec{\xi} \in X} \tilde{a}_{T} (\vec{\xi}, \vec{z})
\end{equation}
where
\begin{equation}
 \tilde{a}_{T} (\vec{\xi}, \vec{z}) = \sum_{\substack{(\vec{\xi}_{1}, \cdots, \vec{\xi}_{n}) \in  X) \\
\vec{\xi}_{1}\circ \cdots \circ \vec{\xi}_{n} = \vec{\xi} \\ T \subset G(\vec{\xi}_{1}, \cdots, \vec{\xi}_{n})}}
\sum_{\substack{(\vec{z}_{1}, \cdots, \vec{z}_{n}) \\ \vec{z}_{1}\circ \cdots \circ \vec{z}_{n} = \vec{z}}} 
 \tilde{a} (\vec{\xi}_{1}, \vec{z}_{1}) \cdots  \tilde{a} (\vec{\xi}_{n}, \vec{z}_{n})
\end{equation}
and
\begin{equation}
\tilde{a} (\vec{\xi}, \vec{z}) = (\pm 1)  \sum_{k=1}^{\infty} \frac{1}{k!}
 \sum_{\substack{T, \text{labelled tree with} \\ \text{vertices} 1,\cdots, k}} a_{T}(\vec{\xi}, \vec{z})
\end{equation}
where  
\begin{equation}
a_{T} (\vec{\xi}, \vec{z}) = \sum_{\substack{(\vec{\xi}_{1}, \cdots, \vec{\xi}_{n}) \in  X) \\
\vec{\xi}_{1}\circ \cdots \circ \vec{\xi}_{k} = \vec{\xi} \\ T \subset G(\vec{\xi}_{1}, \cdots, \vec{\xi}_{k})}}
\sum_{\substack{(\vec{z}_{1}, \cdots, \vec{z}_{k}) \\ \vec{z}_{1}\circ \cdots \circ \vec{z}_{k} = \vec{z}}} 
a (\vec{\xi}_{1}, \vec{z}_{1}) \cdots  a (\vec{\xi}_{k}, \vec{z}_{k})
\end{equation} 
and then use the following lemmas. Lemma 1 expresses the bounds of $|\tilde{a}_{T}|$ and
$|a_{T}|$ in terms of $|\tilde{a}|$ and $|a|$ respectively. 

\textbf{Lemma 1}\cite{BFKT} Let $w_{h_{1}, h_{2}}(\vec{\xi}, \vec{z}) =  e^{\kappa t(\text{supp}(\vec{\xi}, \vec{z}))} h_{1}^{n} h_{2}^{m}$ 
be the weight system with mass $\kappa$ giving weight at least $h_{1}$ to $\psi$ and $h_{2}$ to $\eta$.
Let \textit{T} be a labelled tree with vertices $1, \cdots, n$ and coordination numbers $d_{1}, \cdots, d_{n}$.
Let $a^{\prime}(\vec{\xi}, \vec{z})$ be a coefficient system. Define a new
coefficient system $a^{\prime}_{T}$ by
\begin{equation}
a^{\prime}_{T}(\vec{\xi}, \vec{z}) = \sum_{\substack{(\vec{\xi}_{1}, \cdots, \vec{\xi}_{n}) \in  X) \\
\vec{\xi}_{1}\circ \cdots \circ \vec{\xi}_{n} = \vec{\xi} \\ T \subset G(\vec{\xi}_{1}, \cdots, \vec{\xi}_{n})}}
\sum_{\substack{(\vec{z}_{1}, \cdots, \vec{z}_{n}) \\ \vec{z}_{1}\circ \cdots \circ \vec{z}_{n} = \vec{z}}}
a^{\prime}(\vec{\xi}_{1}, \vec{z}_{1}) \cdots a^{\prime}(\vec{\xi}_{n}, \vec{z}_{n})
\end{equation}
Then
\begin{equation}
|a^{\prime}_{T}|_{w_{1,h}} \leqslant d_{1}! \cdots d_{n}! \hspace{0.05 cm} |a^{\prime}|_{w_{2,h}}^{n}.
\end{equation}

Lemma 2 provides estimates of the sums arising in Lemma 1 assuming that all the norms are less than
$\frac{1}{8}$. Note that the choice of $\frac{1}{8}$ in Lemma 2 is arbitrary.

\textbf{Lemma 2}\cite{BFKT} Let $0 < \epsilon < \frac{1}{8}$. Then
\begin{equation}
\epsilon + \sum_{n=2}^{\infty} \frac{1}{(n-1)!} \sum_{\substack{d_{1}, \cdots, d_{n} \\ d_{1} + \cdots + d_{n} = 2(n-1)}}
 \sum_{\substack{T, \text{labelled tree} \\ \text{with coordination} \\ \text{numbers} \hspace{0.05 cm} d_{1}, \cdots, d_{n} }}
d_{1}! \cdots d_{n}! \hspace{0.05 cm} \epsilon^{n} \leqslant \frac{\epsilon}{1-8\epsilon}. 
\end{equation} 

To estimate norms first rewrite (2.9) 
 \begin{equation}
|b(\vec{z}) - b_{0}|_{w_{h}} \leqslant 
\sum_{n=1}^{\infty} \frac{1}{n!} \sum_{\substack{T, \text{labelled tree with} \\ \text{vertices} 1,\cdots, n}}
\sum_{\vec{\xi} \in X} \hspace{0.05 cm} |\tilde{a}_{T}|_{w_{1,h}}  \nonumber
\end{equation}
Note that from (2.10) and Lemma 1
\begin{equation}
|\tilde{a}_{T}|_{w_{1,h}} \leqslant d_{1}! \cdots d_{n}! \hspace{0.05 cm} |\tilde{a}|_{w_{2,h}}^{n}. \nonumber
\end{equation}
Now using Lemma 2 with $\epsilon = |\tilde{a}|_{w_{2,h}}$ 
\begin{equation}
|b(\vec{z}) - b_{0}|_{w_{h}} \leqslant \frac{|\tilde{a}|_{w_{2,h}}}{1 - 8|\tilde{a}|_{w_{2,h}}}.
\end{equation}
Next rewrite (2.11)
\begin{equation}
|\tilde{a}|_{w_{2,h}} \leqslant \sum_{k=1}^{\infty} \frac{1}{k!} \sum_{\substack{T, \text{labelled tree with} \\ \text{vertices} 1,\cdots, k}}
|a_{T}|_{w_{2,h}}. \nonumber
\end{equation}
Note that from (2.12) and Lemma 1
\begin{equation}
|a_{T}|_{w_{2,h}} \leqslant d_{1}! \cdots d_{k}! \hspace{0.05 cm} |a|_{w_{4,h}}^{k}.       \nonumber
\end{equation}
Now using Lemma 2 with $\epsilon = |a|_{w_{4,h}}$
\begin{equation}
|\tilde{a}|_{w_{2,h}} \leqslant  \frac{|a|_{w_{4,h}}}{1 - 8|a|_{w_{4,h}}}.
\end{equation}
Substituting $|\tilde{a}|_{w_{2,h}}$ from (2.17) in (2.16)
\begin{equation}
|b(\vec{z}) - b_{0}|_{w_{h}} = ||H - H(0)||_{w_{h}}
\leqslant \frac{|a|_{w_{4,h}}}{1 - 16|a|_{w_{4,h}}}.
\end{equation}

\section{Application} 
We give a simple application of how the above theorem can be used to show exponential decay of truncated 
correlation functions in massive Gross-Neveu model on a unit lattice.  
For more sophisticated (bosonic) case we refer reader to our earlier work \cite{G}. That would also be relevant in systems  
containing both fermions and bosons. 

\subsection{The model}
Gross-Neveu model is a theory of massless Dirac fermions having $N$ flavors and interacting via a quartic interaction.
Gross and Neveu \cite{GN} studied the dynamical mass generation in this model through spontaneous symmetry
breaking of the chiral symmetry in two dimensions ($\text{GN}_{2}$). Kopper, Magnen and Rivasseau \cite{KMR} 
studied the mass generation in the large $N$ Euclidean $\text{GN}_{2}$ using an auxiliary bosonic field. 
They proved the existence of a mass gap in the model by showing the exponential decay of two-point function.
Gawedzki and Kupiainen \cite{GK} studied the ultraviolet problem in Euclidean $\text{GN}_{2}$ and showed
the existence of the continuum limit. They relied on the convergence of perturbation series
of effective actions of the model at various momentum scales. 
Feldman, Magnen, Rivasseau and S$\acute{e}$n$\acute{e}$or \cite{FMRS} constructed Schwinger functions
of massive $\text{GN}_{2}$. Our model is massive Gross-Neveu on finite set $\Lambda \subset \mathbb{Z}^{d}$.
We show  exponential decay of truncated correlation function in this model uniform in the volume.
Proving such a decay is an essential part of any rigorous analysis of many-body field theories.

The action is given by
\begin{equation}
\mathcal{A}(\psi, \bar{\psi}) = \sum_{x, \alpha} \sum_{a=1}^{N} 
\bar{\psi}_{\alpha,a}(x) (\mathfrak{D} + m_{f})_{\alpha\beta} \hspace{0.05 cm} \psi_{\beta,a}(x) + V(\psi, \bar{\psi})
\end{equation}
where $a = 1, \cdots, N$ denotes some internal symmetry of fermions, for example, color charge and  
\begin{equation}
V(\psi, \bar{\psi}) = g^{2} \sum_{x} 
\Big(\sum_{\alpha} \sum_{a =1}^{N} \bar{\psi}_{\alpha,a}(x) \psi_{\alpha,a}(x) \Big)^{2}
\end{equation}
$N$ is fixed and the coupling constant $g$ is weak. To study the large $N$ case or limit $N \to \infty$ we have to
take $g$ to depend on $N$. Here we are not interested in that. $\mathfrak{D} = \gamma \cdot \nabla - \frac{1}{2} \Delta$ 
is the Dirac operator.

The generating functional is given by 
\begin{equation}
\mathrm{Z} [\text{J}, \bar{\text{J}}] =  \int_{\Lambda} d\mu_{S}(\psi, \bar{\psi}) \hspace{0.05 cm}
 e^{- V(\psi, \bar{\psi})} e^{ g \langle \bar{\psi}, \text{J}\rangle + g \langle \bar{\text{J}}, \psi \rangle}
\end{equation}
where
\begin{equation}
d\mu_{S}(\psi, \bar{\psi}) = \frac{\prod_{\alpha,a,x} d\bar{\psi}_{\alpha,a}(x) d\psi_{\alpha,a}(x) \hspace{0.05 cm}
e^{- \langle \bar{\psi}, S^{-1} \psi \rangle}}
{\int \prod_{\alpha,a,x} d\bar{\psi}_{\alpha,a}(x) d\psi_{\alpha,a}(x) \hspace{0.05 cm}
e^{- \langle \bar{\psi}, S^{-1} \psi \rangle}}
\end{equation}
$\text{J}, \bar{\text{J}}$ are external sources that are taken to be Grassmann variables and
\begin{equation}
\langle \bar{\psi}, \text{J}\rangle = \sum_{x,a,\alpha}  \bar{\psi}_{\alpha,a}(x)  \text{J}_{\alpha,a}(x) \hspace{1 cm} 
 \langle \bar{\text{J}}, \psi \rangle = \sum_{x,a,\alpha} \bar{\text{J}}_{\alpha,a}(x) \psi_{\alpha,a}(x) 
\end{equation}
$S$ is the massive (and nonlocal) covariance; $S_{\alpha\beta}(x,y) = (\mathfrak{D}(x,y) + m_{f})_{\alpha\beta}^{-1}$
and is known to have a decay
\begin{equation}
|S_{\alpha\beta}(x,y)| < c \hspace{0.05 cm} e^{-m_{f} |x-y|}.
\end{equation} 

\subsection{Power series representation}
The standard procedure in the analysis of the integral (3.3) includes a decoupling expansion for $\mu_{S}$. 
Instead we make the change of variables $\psi \rightarrow S \psi $. Define 
\begin{equation}
\begin{aligned}
V_{1}(\psi, \bar{\psi}, \text{J}, \bar{\text{J}}) &= V(S \psi, \bar{\psi}) 
- g \langle \bar{\psi}, \text{J}\rangle - g \langle \bar{\text{J}}, S \psi \rangle \\
&= g^{2} \sum_{x} \Big(\sum_{\alpha, a}  \bar{\psi}_{\alpha, a}(x) (S \psi)_{\alpha,a} (x) \Big)^{2} 
- g \langle \bar{\psi}, \text{J}\rangle - g \langle \bar{\text{J}}, S \psi \rangle \\
&= g^{2} \sum_{x} 
\Big(\sum_{\alpha, \beta, a, y}  \bar{\psi}_{\alpha, a}(x) S_{\alpha\beta}(x,y) \psi_{\beta,a}(y) \Big)^{2}
- g \langle \bar{\psi}, \text{J}\rangle - g \langle \bar{\text{J}}, S \psi \rangle
\end{aligned}
\end{equation}
Rewrite the generating functional
\begin{equation}
\mathrm{Z} [\text{J}, \bar{\text{J}}] = (\text{det} \hspace{0.05 cm} S)^{-1} \int_{\Lambda} d\mu_{I}(\psi, \bar{\psi}) \hspace{0.05 cm}
e^{- V_{1}(\psi, \bar{\psi}, \text{J}, \bar{\text{J}})}
\end{equation}
Note that
\begin{equation}
\begin{aligned}
V(S \psi, \bar{\psi})  &= g^{2} \sum_{x} 
\Big(\sum_{\alpha, \beta, a, y}  \bar{\psi}_{\alpha, a}(x) S_{\alpha\beta}(x,y) \psi_{\beta,a}(y) \Big)^{2} \\
&= g^{2} \sum_{x} 
 \Big(\sum_{\alpha, \beta, a, y}  \bar{\psi}_{\alpha, a}(x) S_{\alpha\beta}(x,y) \psi_{\beta,a}(y) \Big) 
 \Big(\sum_{\alpha^{\prime}, \beta^{\prime}, a^{\prime}, y^{\prime}} \bar{\psi}_{\alpha^{\prime}, a^{\prime}}(x) 
S_{\alpha^{\prime}\beta^{\prime}}(x,y^{\prime}) \psi_{\beta^{\prime},a^{\prime}}(y^{\prime}) \Big) \\
&= g^{2} \sum_{x} \sum_{\alpha, \beta, a, y} \sum_{\alpha^{\prime}, \beta^{\prime}, a^{\prime}, y^{\prime}}
S_{\alpha\beta}(x,y) S_{\alpha^{\prime}\beta^{\prime}}(x,y^{\prime})
\bar{\psi}_{\alpha, a}(x)  \psi_{\beta,a}(y) \bar{\psi}_{\alpha^{\prime}, a^{\prime}}(x) \psi_{\beta^{\prime},a^{\prime}}(y^{\prime}) \\
&= - g^{2} \sum_{x} \sum_{\alpha, \beta, a, y} \sum_{\alpha^{\prime}, \beta^{\prime}, a^{\prime}, y^{\prime}}
S_{\alpha\beta}(x,y) S_{\alpha^{\prime}\beta^{\prime}}(x,y^{\prime})
\bar{\psi}_{\alpha, a}(x) \bar{\psi}_{\alpha^{\prime}, a^{\prime}}(x) \psi_{\beta,a}(y) \psi_{\beta^{\prime},a^{\prime}}(y^{\prime}) \\
&= \sum_{\substack{x,y,y^{\prime} \\ \alpha,\beta,\alpha^{\prime},\beta^{\prime} }} \sum_{a,a^{\prime}}
a_{ \alpha,\beta,\alpha^{\prime},\beta^{\prime} } (x,y,y^{\prime}) \hspace{0.05 cm}
\bar{\psi}_{\alpha, a}(x) \bar{\psi}_{\alpha^{\prime}, a^{\prime}}(x) \psi_{\beta,a}(y) \psi_{\beta^{\prime},a^{\prime}}(y^{\prime}) 
\end{aligned}
\end{equation}
where
\begin{equation}
a_{ \alpha,\beta,\alpha^{\prime},\beta^{\prime} } (x,y,y^{\prime}) =
- g^{2} S_{\alpha\beta}(x,y) S_{\alpha^{\prime}\beta^{\prime}}(x,y^{\prime})
\end{equation}
which is neither symmetric nor antisymmetric. Let $t(x,y,y^{\prime})$ be the length of the shortest tree
joining $(x,y,y^{\prime})$. Then from (3.6) 
\begin{equation}
|a_{ \alpha,\beta,\alpha^{\prime},\beta^{\prime}} (x,y,y^{\prime})| \leqslant c \hspace{0.05 cm} g^{2} \hspace{0.05 cm} 
e^{-m_{f} |x-y|} e^{-m_{f} |x-y^{\prime}|} \leqslant  c \hspace{0.05 cm} g^{2}  \hspace{0.05 cm}
e^{-m_{f} t(x,y,y^{\prime})}.
\end{equation}
The source term
\begin{equation}
\begin{aligned}
g \langle \bar{\text{J}}, S \psi \rangle = 
g \sum_{x,a,\alpha} \bar{\text{J}}_{\alpha,a}(x) (S \psi)_{\alpha,a}(x) 
&=  g \sum_{x,y,a,\alpha,\beta} \bar{\text{J}}_{\alpha,a}(x) S_{\alpha \beta}(x,y) \psi_{\beta,a}(y) \\
&= \sum_{x,y,a,\alpha,\beta} a_{\alpha, \beta}(x,y) \bar{\text{J}}_{\alpha,a}(x) \psi_{\beta,a}(y) 
\end{aligned}
\end{equation}
where $ a_{\alpha, \beta}(x,y) = g \hspace{0.05 cm} S_{\alpha \beta}(x,y)$ and 
$|a_{\alpha, \beta}(x,y)| \leqslant c \hspace{0.05 cm} g \hspace{0.05 cm} e^{-m_{f}|x-y|}$.
 
\textbf{Norm}. Let $w_{h_{1}, h_{2}}(\vec{\xi}, \vec{z}) =  e^{\kappa t(\text{supp}(\vec{\xi}, \vec{z}))} h_{1}^{n} h_{2}^{m}$ 
be the weight system with mass $\kappa$ giving weight at least $h_{1}$ to $\psi$ and $h_{2}$ to 
$\bar{\text{J}}, \text{J}$. Set $h_{1} = 4$ and $h_{2} = 1$. Then 
$||V||_{w_{4,1}} = |a_{ \alpha,\beta,\alpha^{\prime},\beta^{\prime} }|_{w_{4,1}}$.
Let $\kappa < m_{f}$ and $n=4, m=0$ in (1.17) and let
$a_{\alpha,\beta,\alpha^{\prime},\beta^{\prime}}(x,y,y^{\prime})$ denote antisymmetrization of $a$ then
\begin{equation}
\begin{aligned}
||V||_{w_{4,1}} &\leqslant \max\limits_{z \in X}  \hspace{0.05 cm} \max\limits_{1\leqslant i \leqslant 4} 
\sum_{\substack{x,y,y^{\prime}  \in X \\ x\hspace{0.05 cm}\text{or}\hspace{0.05 cm}  y \hspace{0.05 cm} \text{or}\hspace{0.05 cm}  y^{\prime} = z}}  
\sum_{\substack{\alpha,\beta,\alpha^{\prime},\beta^{\prime}  \\ a,a^{\prime}}}  
e^{\kappa t(x,y,y^{\prime})} \hspace{0.05 cm} h_{1}^{4} |a_{\alpha,\beta,\alpha^{\prime},\beta^{\prime}}(x,y,y^{\prime})|_{w_{4,1}} \\
&\leqslant \max\limits_{z \in X}  \hspace{0.05 cm} \max\limits_{1\leqslant i \leqslant 4} 
\sum_{\substack{x,y,y^{\prime}  \in X \\ x\hspace{0.05 cm}\text{or}\hspace{0.05 cm}  y\hspace{0.05 cm}\text{or}\hspace{0.05 cm}  y^{\prime} = z}}  
\sum_{\substack{\alpha,\beta,\alpha^{\prime},\beta^{\prime}  \\ a,a^{\prime}}}  
e^{\kappa t(x,y,y^{\prime})} \hspace{0.05 cm}  
c \hspace{0.05 cm} 4^{4} \hspace{0.05 cm} g^{2} \hspace{0.05 cm} e^{-m_{f} t(x,y,y^{\prime})} \\
&\leqslant \max\limits_{z \in X}  \hspace{0.05 cm} \max\limits_{1\leqslant i \leqslant 4} 
\sum_{\substack{x,y,y^{\prime} \in X \\ x\hspace{0.05 cm} \text{or}\hspace{0.05 cm} y\hspace{0.05 cm}\text{or}\hspace{0.05 cm}y^{\prime} = z}}  
d^{4} N^{2} c \hspace{0.05 cm} (16\hspace{0.05 cm} g)^{2}  \hspace{0.05 cm} e^{-(m_{f} - \kappa) t(x,y,y^{\prime})} \\
 &\leqslant  c \hspace{0.05 cm} (16\hspace{0.05 cm} g)^{2} N^{2} d^{4}.
\end{aligned}
\end{equation}
For the source term set $h_{1} = 4$ and $h_{2} = 1$ and let $\kappa < m_{f}$ with $n=1, m=1$ in (1.17)
\begin{equation}
\begin{aligned}
g||\langle \bar{\text{J}}, S \psi \rangle||_{w_{4,1}} &\leqslant \max\limits_{z \in X}  \hspace{0.05 cm}
\sum_{\substack{x,y \in X \\ x\hspace{0.05 cm}\text{or}\hspace{0.05 cm}  y = z}}
\sum_{\alpha,\beta, a} e^{\kappa |x-y|} \hspace{0.05 cm}  
c \hspace{0.05 cm} 4 \hspace{0.05 cm}  g \hspace{0.05 cm} e^{-m_{f}|x-y|} \\
&\leqslant \max\limits_{z \in X}  \hspace{0.05 cm}
\sum_{\substack{x,y \in X \\ x\hspace{0.05 cm}\text{or}\hspace{0.05 cm}  y = z}}
d^{2} N \hspace{0.05 cm} c \hspace{0.05 cm} (4 \hspace{0.05 cm} g) \hspace{0.05 cm} e^{-(m_{f} - \kappa) |x-y|} \\
&\leqslant c \hspace{0.05 cm} (4\hspace{0.05 cm} g) N \hspace{0.05 cm} d^{2}.
\end{aligned}
\end{equation}
Then
\begin{equation}
\begin{aligned}
||V_{1}||_{w_{4,1}} &\leqslant ||V||_{w_{4,1}} + 2 g ||\langle \bar{\text{J}}, S \psi \rangle||_{w_{4,1}} \\
&\leqslant c \hspace{0.05 cm} (4 \hspace{0.05 cm} g)  N \hspace{0.05 cm} d^{2}.
\end{aligned}
\end{equation}

Let  $z$ stands for $(y, \alpha, a, \omega)$ with $\omega = (0,1)$. Define
 \begin{equation}
 \text{J}(z) = \begin{cases}
 \text{J}_{\alpha,a}(y)      & \text{if} \hspace{0.4 cm} z = (y, \alpha, a, 0) \\
\bar{\text{J}}_{\alpha,a}(y)   & \text{if} \hspace{0.4 cm} z = (y, \alpha, a, 1)
\end{cases}
\end{equation} 
The theorem implies that given a power series of $V_{1}(\psi, \bar{\psi}, \text{J}, \bar{\text{J}})$ 
with series coefficients as above there exists a coefficient system $b(\vec{z})$ having same form as 
series coefficients of $V_{1}(\psi, \bar{\psi}, \text{J}, \bar{\text{J}})$ (i.e. $b(\vec{z})$ are neither symmetric
nor antisymmetric and have tree decay property) such that
\begin{equation}
\begin{aligned}
\text{log} \hspace{0.1 cm} \mathrm{Z} [\text{J}] &= b_{0} + 
\sum_{z_{1},z_{2}}  \text{J}(z_{1}) \hspace{0.05 cm} b(z_{1}, z_{2}) \hspace{0.05 cm} \text{J}(z_{2}) +
\sum_{z_{1}, z_{2}, z_{3}, z_{4}} \text{J}(z_{1}) \text{J}(z_{2}) \hspace{0.05 cm} 
b(z_{1}, z_{2}, z_{3}, z_{4}) \hspace{0.05 cm} \text{J}(z_{3}) \text{J}(z_{4}) +  \cdots \\
&=  \text{log} \hspace{0.1 cm} \mathrm{Z} [0] + 
\sum_{m \geqslant 1} \sum_{z_{1}, \cdots, z_{2m}}
 \text{J}(z_{1}) \cdots \text{J}(z_{m}) \hspace{0.05 cm} b(z_{1}, \cdots, z_{2m}) 
\hspace{0.05 cm} \text{J}(z_{m+1}) \cdots \text{J}(z_{2m}) 
\end{aligned}
\end{equation}
Denote $\text{log} \hspace{0.1 cm} \mathrm{Z} [\text{J}] = H(\text{J})$ then 
if $||V_{1}||_{w_{4,1}}  < \frac{1}{16}$
\begin{equation}
||H - H(0)||_{w_{1}} 
\leqslant \frac{||V_{1}||_{w_{4,1}}}{1 - 16 ||V_{1}||_{w_{4,1}}}
\leqslant  c \hspace{0.05 cm} (4\hspace{0.05 cm} g)  N \hspace{0.05 cm} d^{2}.
\end{equation}
The norm $||H - H(0)||_{w_{1}}$ is actually better than $\mathcal{O}(g)$. To see this note that
\begin{equation}
\begin{aligned}
\mathrm{Z} [\text{J}, \bar{\text{J}}] &= (\text{det} \hspace{0.05 cm} S)^{-1} \int_{\Lambda} d\mu_{I}(\psi, \bar{\psi}) \hspace{0.05 cm}
e^{- V_{1}(\psi, \bar{\psi}, \text{J}, \bar{\text{J}})} \\
&= (\text{det} \hspace{0.05 cm} S)^{-1} \int_{\Lambda} d\mu_{I}(\psi, \bar{\psi}) \big[1+ \sum_{n=1}^{\infty} \frac{1}{n!} \hspace{0.05 cm}
a(\xi_{1}, \cdots, \xi_{n}) \psi(\xi_{1}) \cdots \psi(\xi_{n})\big] \\
& \hspace{2 cm} \big[1 + \sum_{m=2}^{\infty} \frac{1}{m!} \hspace{0.05 cm}
(g \langle \bar{\psi}, \text{J}\rangle + g \langle \bar{\text{J}}, S \psi \rangle)^{m} \big] \nonumber
\end{aligned}
\end{equation}
The first integral is even in $\psi$ which forces the second integral to start from $m=2$ 
to keep $\psi$ even (integral with $m=1$ term would be 0). Therefore, since every term in the expansion
is $\mathcal{O}(g^{2})$ the norm $||H - H(0)||_{w_{1}} $ is  
\begin{equation}
||H - H(0)||_{w_{1}}  \leqslant c \hspace{0.05 cm} (16\hspace{0.05 cm} g^{2}) N^{2} \hspace{0.05 cm} d^{4}.
\end{equation}

\textbf{Estimate of $b(\vec{z})$}. Rewrite (1.18) as 
\begin{equation}
|b(z_{1}, \cdots, z_{2m})|  \leqslant h^{- 2m} e^{- \kappa t(z_{1}, \cdots, z_{2m})}
||H - H(0)||_{w_{h}}  \nonumber
\end{equation}
Set $h =  1$ and $\kappa < m_{f}$. Then
\begin{equation}
\begin{aligned}
|b(z_{1}, \cdots, z_{2m})|  &\leqslant   e^{- \kappa t(z_{1}, \cdots, z_{2m})}
||H - H(0)||_{w_{1}}  \\
&\leqslant  c \hspace{0.05 cm} (16\hspace{0.05 cm} g^{2})  N^{2} \hspace{0.05 cm} d^{4} \hspace{0.05 cm} 
e^{- \kappa t(z_{1}, \cdots, z_{2m})}.
\end{aligned}
\end{equation}

\subsection{Correlation functions}
The truncated correlation function is defined as
\begin{equation}
\begin{aligned}
\langle \bar{\psi}_{\beta, b}(y_{1}) \psi_{\alpha, a}(y_{2}) \rangle - 
\langle  \bar{\psi}_{\beta, b}(y_{1})\rangle \langle \psi_{\alpha, a}(y_{2}) \rangle
&= \frac{1}{g^{2}} \left. \frac{\delta}{\delta \bar{\text{J}}_{\alpha,a}(y_{1})} \text{log}\hspace{0.05 cm} 
\mathrm{Z} [\text{J},\bar{\text{J}}] \frac{\delta}{\delta \text{J}_{\beta,b}(y_{2})} \right\rvert_{\text{J} = \bar{\text{J}} = 0} \\
\end{aligned}
\end{equation}
We have written a power series of $\text{log} \hspace{0.1 cm} \mathrm{Z} [\text{J}]$
\begin{equation}
\text{log} \hspace{0.1 cm} \mathrm{Z} [\text{J}] = b_{0} + 
\sum_{z_{1},z_{2}}  \text{J}(z_{1}) \hspace{0.05 cm} b(z_{1}, z_{2}) \hspace{0.05 cm} \text{J}(z_{2}) +
 \text{higher order terms}     \nonumber
\end{equation}
Note that
\begin{equation}
\begin{aligned}
\sum_{z_{1},z_{2}}  \text{J}(z_{1}) \hspace{0.05 cm} b(z_{1}, z_{2}) \hspace{0.05 cm} \text{J}(z_{2})
= \sum_{y_{1},y_{2}} & \sum_{\alpha,\beta} \sum_{a,b}  \bar{\text{J}}_{\alpha,a}(y_{1}) \hspace{0.05 cm} 
b_{\alpha,\beta,a,b} (y_{1},y_{2}) \hspace{0.05 cm} \text{J}_{\beta,b}(y_{2})  \\
& + \sum_{y_{1},y_{2}} \sum_{\alpha,\beta} \sum_{a,b} \text{J}_{\beta,b}(y_{1}) \hspace{0.05 cm} 
b_{\beta,\alpha,b,a} (y_{2},y_{1}) \hspace{0.05 cm}  \bar{\text{J}}_{\alpha,a}(y_{2})
\end{aligned}
\end{equation}
Thus
\begin{equation}
\left. \frac{\delta}{\delta \bar{\text{J}}_{\alpha,a}(y_{1})} \text{log}\hspace{0.05 cm} 
\mathrm{Z} [\text{J},\bar{\text{J}}] \frac{\delta}{\delta \text{J}_{\beta,b}(y_{2})} \right\rvert_{\text{J} = \bar{\text{J}} = 0} 
=  b_{\alpha,\beta,a,b} (y_{1},y_{2}) -  b_{\beta,\alpha,b,a} (y_{2},y_{1}).
\end{equation}
Since $\text{J} = \bar{\text{J}} = 0$ only one term survives; sum over $\alpha,\beta,a,b, y_{1}, y_{2}$ collapses to 
a single term. Therefore, 
\begin{equation}
\begin{aligned}
|\langle \bar{\psi}_{\beta, b}(y_{1}) \psi_{\alpha, a}(y_{2}) \rangle - 
& \langle  \bar{\psi}_{\beta, b}(y_{1})\rangle \langle \psi_{\alpha, a}(y_{2}) \rangle| 
\leqslant \frac{1}{g^{2}} |b_{\alpha,\beta,a,b} (y_{1},y_{2}) - b_{\beta,\alpha,b,a} (y_{2},y_{1})| \\
&\leqslant \frac{c}{g^{2}} \hspace{0.05 cm} (16\hspace{0.05 cm}g^{2}) N^{2} \hspace{0.05 cm} d^{4} \hspace{0.05 cm} 
 e^{- \kappa |y_{1} - y_{2}|} \\
&\leqslant c \hspace{0.05 cm} 16 \hspace{0.05 cm} N^{2} \hspace{0.05 cm} d^{4}\hspace{0.05 cm} 
 e^{- \kappa |y_{1} - y_{2}|}.
\end{aligned}
\end{equation}

\textbf{Theorem 2} Given a Gross-Neveu model (3.1-3.3) on a finite lattice in $d$ dimensions.
Let the coupling constant $g$ be small enough so that $||V_{1}||_{w_{4,1}} < 1/16$ and $N$ remains fixed. 
Then there exists a constant $0 < \kappa < m_{f}$ independent of the lattice size such that
\begin{equation}
|\langle \bar{\psi}_{\beta, b}(y_{1}) \psi_{\alpha, a}(y_{2}) \rangle - 
 \langle  \bar{\psi}_{\beta, b}(y_{1})\rangle \langle \psi_{\alpha, a}(y_{2}) \rangle| 
\leqslant   c \hspace{0.05 cm}16 \hspace{0.05 cm} N^{2} \hspace{0.05 cm} d^{4} \hspace{0.05 cm} 
 e^{- \kappa |y_{1} - y_{2}|}.
\end{equation}

\textbf{Remark 4}. A similar result is established in non relativistic many-body fermionic lattice system by 
De Roeck and Salmhofer \cite{DS}. They show that if the Hamiltonian of a free theory has a gap and if the interaction terms 
are weak and have a local structure then the Hamiltonian of the interacting theory also has a gap (smaller than that of the free theory).
This is nicely captured in the idea of our expansion which is more adapted to relativistic systems. 
The decay nature of the free propagator is directly responsible for the decay of truncated correlation functions 
(hence the gap) with a reduced decay rate.

\textbf{Acknowledgement}. I would like to thank Jonathan Dimock for helpful comments and
suggestions.

\end{document}